\documentclass[final,5p,times,twocolumn]{elsarticle}
\biboptions{sort&compress}

\usepackage{amsmath,amssymb}
\usepackage{graphicx} % FOR INCLUDING FIGURES
\usepackage{subfigure} % FOR SUBFIGURES

\journal{Physica D}

\begin{document}

\begin{frontmatter}
\title{Aggressive shadowing of a low-dimensional model of atmospheric dynamics}
\author[UVM]{Ross M. Lieb-Lappen\corref{cor1}}
\ead{rlieblappen@gmail.com}

\author[UVM,VACC,CSC]{Christopher M. Danforth}
\ead{chris.danforth@uvm.edu}

\cortext[cor1]{Corresponding author}

\address[UVM]{Department of Mathematics and Statistics, University of Vermont, Burlington, VT}
\address[VACC]{Vermont Advanced Computing Center, University of Vermont, Burlington, VT}
\address[CSC]{Complex Systems Center, University of Vermont, Burlington, VT}

\begin{abstract}
Predictions of the future state of the Earth's atmosphere suffer from the consequences of chaos: numerical weather forecast models quickly diverge from observations as uncertainty in the initial state is amplified by nonlinearity.  One measure of the utility of a forecast is its \textit{shadowing time}, informally given by the period of time for which the forecast is a reasonable description of reality.  The present work uses the Lorenz $'96$ coupled system, a simplified nonlinear model of atmospheric dynamics, to extend a recently developed technique for lengthening the shadowing time of a dynamical system.  Ensemble forecasting is used to make forecasts with and without \textit{inflation}, a method whereby the ensemble is regularly expanded artificially along dimensions whose uncertainty is contracting.  The first goal of this work is to compare model forecasts, with and without inflation, to a true trajectory created by integrating a modified version of the same model.  The second goal is to establish whether inflation can increase the maximum shadowing time for a single optimal member of the ensemble.  In the second experiment the true trajectory is known a priori, and only the closest ensemble members are retained at each time step, a technique known as \textit{stalking}.  Finally, a \textit{targeted inflation} is introduced to both techniques to reduce the number of instances in which inflation occurs in directions likely to be incommensurate with the true trajectory.  Results varied for inflation, with success dependent upon the experimental design parameters (e.g. size of state space, inflation amount).  However, a more targeted inflation successfully reduced the number of forecast degradations without significantly reducing the number of forecast improvements.  Utilized appropriately, inflation has the potential to improve predictions of the future state of atmospheric phenomena, as well as other physical systems.  
\end{abstract}
 
\begin{keyword}
shadowing \sep stalking \sep ensemble forecasting \sep inflation
\end{keyword}

\end{frontmatter}

\section{Introduction}
Society is often dependent upon the ability of scientists to accurately forecast the future state of chaotic physical systems, such as the weather.  In extreme cases, atmospheric scientists are asked to anticipate natural disasters with ample time for the public to adequately prepare.  To predict the likelihood of atmospheric phenomena, meteorologists estimate Earth's future atmospheric state using sophisticated computer simulations.  In general, these mathematical models struggle to predict the specific behavior of chaotic physical systems due to $(1)$ uncertainty in the initial state, $(2)$ chaos, and $(3)$ model error \cite{Leith78}. \par 

The initial state is an estimate of the condition of the atmosphere at the beginning of a forecast.  Although weather monitoring devices observe much of the planet, for many areas (e.g. southern hemisphere oceans, the upper atmosphere) the measurements are spatially sparse and typically three orders of magnitude less in number than the degrees of freedom in a global weather model.  Therefore, meteorologists must combine observations with prior forecasts to estimate initial conditions in a process called \textit{data assimilation}.  \par 

Since perfect knowledge of atmospheric tendencies is unachievable, a great deal of recent research has focused on data assimilation, the process by which observations are combined with model predictions to give the best possible initial state, or \textit{analysis} \cite{Whitaker08, Hamill05, Hamill06, Hunt07, Ott04}.  It is the analysis that is typically used as a proxy for the true state of the atmosphere at any time in the past.  Although the analysis has inherent uncertainty from lack of perfect observations and an imperfect model, the modeler's goal is to create a forecast from this given state that remains reasonable (i.e. close to observations) for the longest duration of time.  In the context of weather predictions, forecasters strive to be accurate for two weeks, the limit imposed by chaos.  Today, five-day forecasts are as good as three-day forecasts were $30$ years ago \cite{Kalnay03}.  \par 

Despite these efforts, the mischaracterization of Earth's atmospheric state introduces uncertainty.  In addition, the atmosphere is a chaotic system, causing these small uncertainties in the initial state to grow exponentially in time.  Finally, since the model itself is not a perfect representation of reality, inaccurate changes forecast by the model, namely model error, compound the uncertainty and lead forecasts to diverge quickly from observations.  As Lorenz noted in $1965$, the limit of predictability of the atmosphere is about two weeks, even with nearly perfect knowledge of the current state \cite{Lorenz65}.  \par 

To account for initial state uncertainty, an accepted technique is to use \textit{ensemble forecasting}, where a collection of perturbations to the best guess are chosen randomly from a distribution that reflects the measurement uncertainty and system dynamics near the given initial state \cite{Toth93}.  Each ensemble member is forecast forward in time, yielding a collection of final states.  This collection roughly represents the probability distribution of the model's forecast with associated uncertainty.  For example, if $60\%$ of the ensemble members predict rain, the forecaster assigns a $60\%$ chance of rain. Given the limitations, the modeler's goal can be stated as trying to keep some ensemble members close to the observed truth for as long as possible, analogous to finding a shadowing trajectory for Earth's atmosphere.\par

Traditional mathematical shadowing theory was developed for hyperbolic systems, based on the Shadowing Lemma of Anosov \cite{Anosov67} and formalized by Bowen \cite{Bowen75}.  Given a pseudo-trajectory of a model (i.e. one very close to an actual one-step trajectory), this lemma establishes the existence of a true trajectory that remains close for an arbitrary period of time.  Later research has extended the lemma for a wide variety of hyperbolic systems (e.g. \cite{Robinson77, Meyer87,Sauer91}).  For these systems, the number of expanding directions remains constant, i.e. the number of Lyapunov exponents greater than zero is constant throughout the state-space, and small perturbations in stable directions decay exponentially in time.  However, most physical systems (e.g. Earth's atmosphere) are non-hyperbolic, i.e. the number of positive Lyapunov exponents fluctuates along a trajectory through state space.  For these systems, there does not exist a trajectory that shadows the truth for an arbitrarily long time \cite{Judd08, Smith99, Orrell01}.  More recent work has been focused on finding shadowing trajectories for low-dimensional non-hyperbolic systems (e.g. general ordinary differential equations \cite{Coomes95, Hammel87}, the driven pendulum \cite{Grebogi90}, the H\'enon map \cite{Hammel88}).  \par 

For the purposes of the present research, the time period for which a particular forecast is an accurate representation of reality is referred to as the \textit{shadowing time}.  Danforth and Yorke \cite{Danforth06} proposed an aggressive approach to shadowing called \textit{stalking} to increase the shadowing time for a given forecast.  For a system with $n$-degrees of freedom, an $n$-dimensional sphere was used to encompass the initial ensemble members.  Throughout the length of the forecast, these members experience expansion away from each other in some dimensions and contraction towards each other in others, depending on the local finite-time Lyapunov exponent in each dimension.  The ensemble forecast can be approximated by an $n$-dimensional ellipsoid for a short time. \par 

The idea of stalking is to artificially impose some additional uncertainty in the contracting directions (i.e. those with a negative local finite-time Lyapunov exponent) at periodic intervals throughout the forecast.  This is accomplished by inflating the ellipsoid along axes parallel to the contracting dimensions.   Thus, if the true state of the system happens to suddenly expand along a previously contracting direction, as happens in systems exhibiting \textit{unstable dimension variability} \cite{Dawson94}, some ensemble members will remain relatively close to the true state.  Stalking is not currently used in weather forecasting, but an alternative version called \textit{variance inflation} is used in ensemble-based data assimilation to ensure the state estimation algorithm does not put too much faith in the model forecasts and ignore observations when creating the analysis \cite{Li09}.  \par 

For the purpose of this paper, the terms ``shadowing" and ``stalking" are reserved for experiments where the truth is known a priori, and the goal is to identify an optimal ensemble member (section V).  Using the technique described above to create forecasts, without prior knowledge of the truth, is here denoted forecasting with and without inflation (section IV).  In section II, we present the non-hyperbolic '96 Lorenz system \cite{Lorenz96} and the model, and examine the effect of changing basic parameters.  In section III, the process of creating the initial ensemble and making forecasts is explained.  In section IV, the inflation technique is introduced, and forecasting results (with and without inflation) are presented.  In section V, we present the results from stalking experiments.  In section VI, we use targeted inflation to isolate the instances in which inflation is most likely to improve a forecast.  Finally, the findings are discussed with respect to operational weather prediction in section VII.\par         

\section{The Lorenz 1996 System} 
In numerical weather prediction (NWP), the true evolution of the Earth's atmosphere will ideally  be a plausible member of an ensemble forecast \cite{Toth93}.  In our context, ensemble forecasting is used to model the trajectory $\textbf{z}^a$, the ``truth," of some meteorological quantity (e.g. temperature, pressure).  In the present study, both the truth and the forecast were created using versions of a simplified nonlinear model given by Lorenz (1996) to represent the atmospheric behavior at $I$ equally spaced locations on a given latitude circle \cite{Lorenz96}.  This system has been used to represent weather related dynamics in many previous studies (e.g. \cite{Wilks05, Orrell02, Danforth08}).  The governing first-order differential equations are given by \cite{Lorenz96}:

\begin{subequations}
\begin{eqnarray}
\dfrac{dx_{i}}{dt}&=&x_{i-1}(x_{i+1}-x_{i-2})-x_{i}+F \nonumber \\*
& & -\dfrac{hc}{b}\sum_{j=J(i-1)+1}^{iJ}y_{j}, \label{slow} \\
\dfrac{dy_{j}}{dt}&=&-cby_{j+1}(y_{j+2}-y_{j-1})-cy_{j} \nonumber \\*
& & +\dfrac{hc}{b}x_{\text{floor}[(j-1)/J]+1} \label{fast}
\end{eqnarray}
\end{subequations}
for $i=1,2,\dots,I$, $j=1,2,\dots,JI$, and $n=(J+1)I$.  Although $n$ represents the full state space size, we are primarily interested in the values of the slow variables $x_{i}$.  Thus, for the purposes of this study we consider $I$ to be the dimensionality of the system.\par  

\begin{figure}
\includegraphics{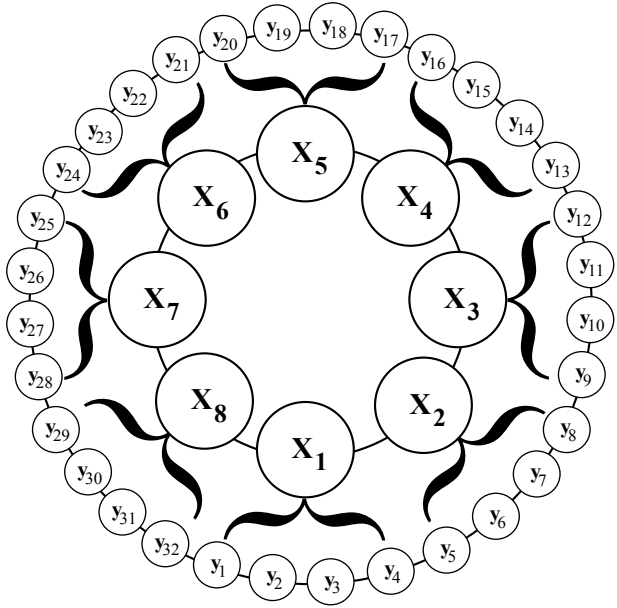}
\caption{\label{schematic}Model Schematic \cite{Wilks05}.  The $I=8$ slow variables $x_{i}$ can be thought of as locations along a latitude circle.  Each variable has $J=4$ corresponding $y_{j}$ fast variables for a total of $n=40$ degrees of freedom.}
\end{figure}

The values $x_{i}$ represent slowly changing meteorological quantities whose dynamics are described by Eq.~\eqref{slow}.  Since the set of $x_{i}$ corresponds to locations along a single latitude circle, the subscripts $i$ and $j$ are defined to be in a cyclic chain.  That is, we define $x_{-1}=x_{I-1}$, $x_{0}=x_{I}$, and $x_{1}=x_{I+1}$, and similarly for $j$.  Each $x_{i}$ is then coupled to $J$ quickly changing, small amplitude variables whose behavior is governed by Eq.~\eqref{fast}.  For our experiments, we set $I=4, 5, 6$ and $J=16$ yielding $n=68, 85,$  and $102$ variables.  A schematic is shown in Fig.~\ref{schematic} where eight slow variables ($x_{i}$) are each coupled to four fast variables ($y_{j}$), reflecting a total of $n=40$ degrees of freedom.  Note that the dynamics of each $x_{i}$ are dictated by neighboring $x$ variables and the corresponding set of coupled $y_{j}$ variables.  \par

As a low-dimensional model of atmospheric dynamics, this system is ideal for several reasons; primary among them being the ability to tune the relative levels of nonlinearity, coupling of timescales, and spatial degrees of freedom.  The nonlinear terms in Eq.~\eqref{slow} are meant to represent advection, and conserve the total energy of the system.  The linear term signifies a loss of energy either through mechanical or thermal dissipation.  External forcing $F$ is then added to prevent the total energy from decaying completely.  For all experiments we set $F=14$.  Consistent with the literature, we set $c=10$ and $b=10$, which forces the fast variables to oscillate $10$ times quicker than the slow variables \cite{Wilks05, Danforth08, Orrell01}.  The coupling parameter is set to a value of $h=1$ for the system truth, and $h=0.5$ for the model.  Note that one time unit corresponds to five days, the dissipative decay time \cite{Lorenz98}.  \par 

Integration of the differential equations is completed using the fourth order Runge-Kutta method with a time step of $0.01$.  Rigorous shadowing attempts could be made using far more advanced methods of integration, with much smaller time steps \cite{Hayes10}.  However, for the purpose of this study of short forecasts, the difference is negligible.  \par

\subsection{Dynamics}
\begin{figure}
\subfigure[\label{wave0}Days $0-5$]{\includegraphics{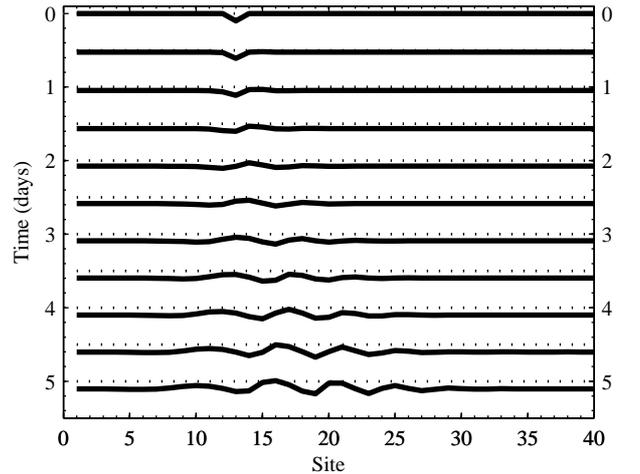}} 
\\
\subfigure[\label{wave50}Days $50-55$]{\includegraphics{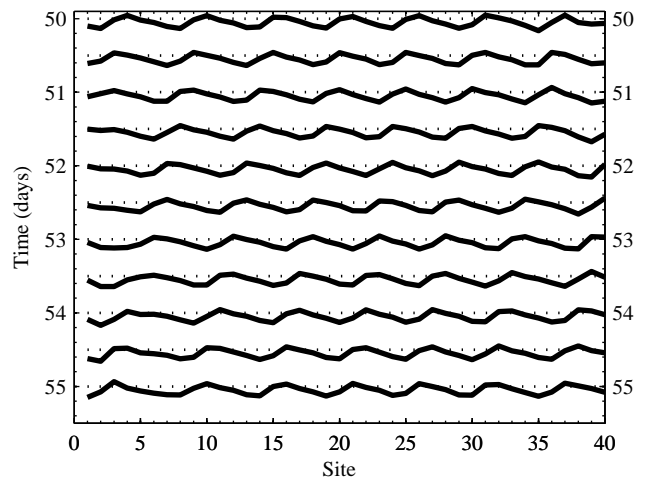}} 
\caption{\label{wave}A time series of the system shows the effect of a five-unit perturbation.  Panel (a) shows days $0-5$ and panel (b) shows days $50-55$.  Perturbations appear to move primarily to the east (right).  After five days, perturbation effects can be observed at over half of all the locations ($I=40, J=16$).}
\end{figure}

The system dynamics can be observed by integrating Eqs.~\eqref{slow} and \eqref{fast} with $h=1$, hereafter referred to as the \textit{system}.  In Fig.~\ref{wave}, a time series for a particular longitudinal profile ($I=40$, $J=16$) is shown after variable $x_{13}$ is initially perturbed by five units.  Profiles are recorded at $12$-h intervals for days $0-5$ (panel a) and days $50-55$ (panel b).  Advection is apparent as the energy from the perturbation is observed halfway around the latitudinal circle by day five.  Lorenz and Emanuel \cite{Lorenz98} calculated a perturbation growth rate (doubling time) of approximately two days for this model, which agrees with trends in larger atmospheric models where errors double in roughly two days \cite{Simmons95}.  However, growth rates over limited time intervals as in Fig.~\ref{wave50} can differ greatly depending on $I$ and $J$. \par

Adjusting the parameters of the system, and examining the time series at a single latitudinal site, the interdependence of the slow and fast variable coupling is observed.  For example, with $I=6$ and $J=16$,  a ``regular" oscillatory pattern occurs as an energy equilibrium is achieved between external forcing and dissipation.  In this work, we will use the term ``regular" to refer to these quasi-periodic patterns.  However, as the significance of the fast variables is reduced by lowering the coupling parameter $h$, the time series becomes more complex.  Less energy is able to dissipate from slow variables to fast, leading to greater irregularity in $x_{i}$. \par 

\begin{figure*}
\subfigure[$I=4, J=8$]{\includegraphics{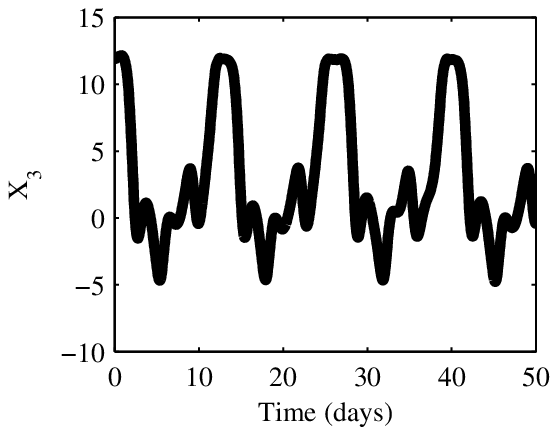}}
\subfigure[$I=4, J=16$]{\includegraphics{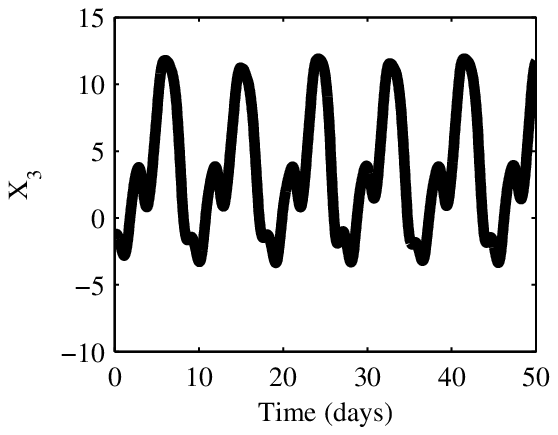}}
\subfigure[$I=4, J=40$]{\includegraphics{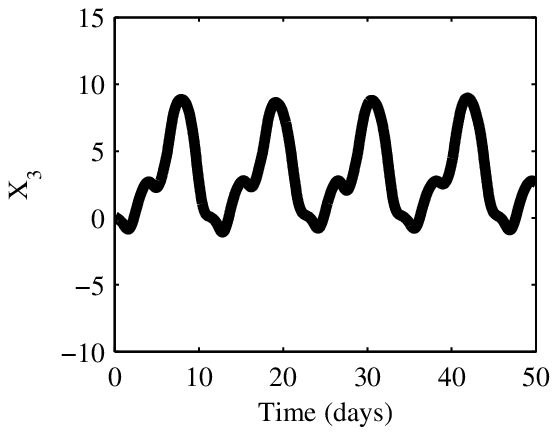}}
\\
\subfigure[$I=6, J=8$]{\includegraphics{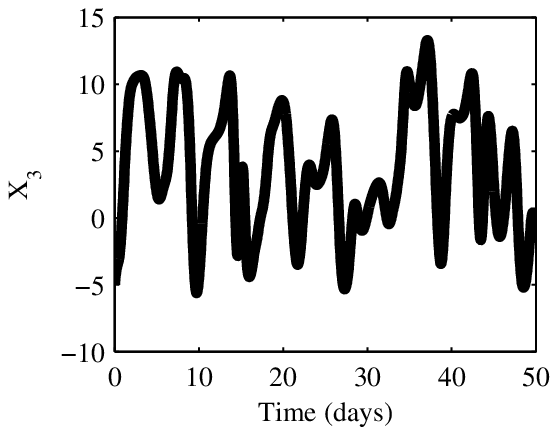}}
\subfigure[$I=6, J=16$]{\includegraphics{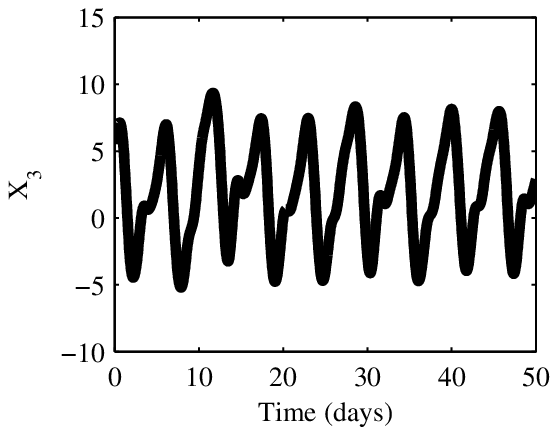}}
\subfigure[$I=6, J=40$]{\includegraphics{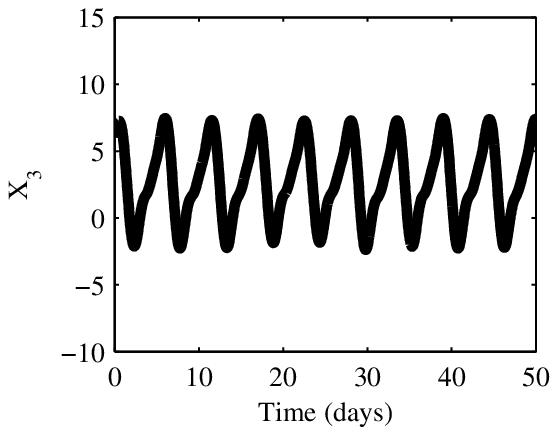}}
\\
\subfigure[$I=8, J=8$]{\includegraphics{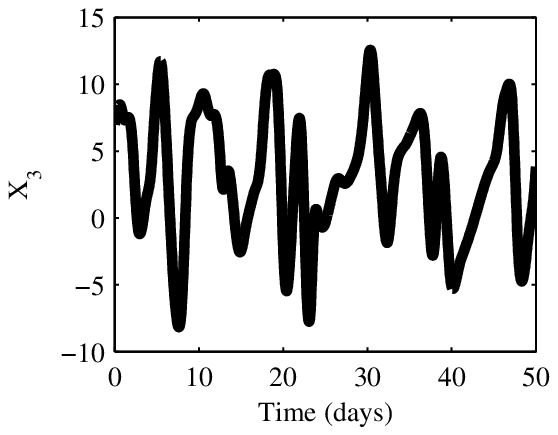}}
\subfigure[$I=8, J=16$]{\includegraphics{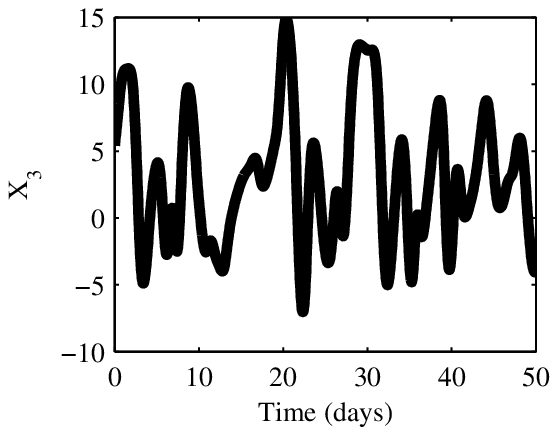}}
\subfigure[$I=8, J=40$]{\includegraphics{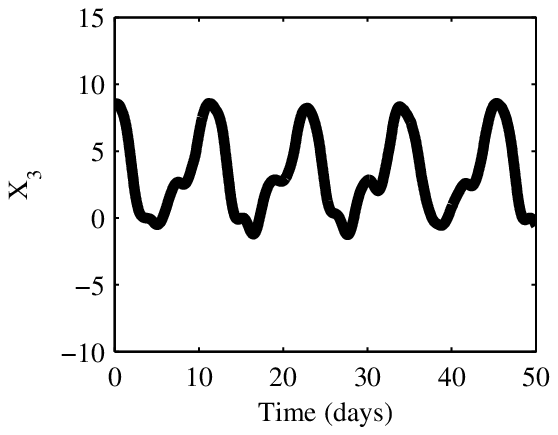}}
\caption{\label{changemn}  A $50$ day time series for $x_{3}$ illustrates the dependence of the system stability on the number of slow $(I)$ and fast $(J)$ variables.  The number of slow variables varies from $I=4$ (a-c) to $I=6$ (d-f) to $I=8$ (g-i).  The fast variables vary from $J=8$ (left column) to $J=16$ (middle column) to $J=40$ (right column).  Generally, increasing $J$ leads to increased regularity, while changes in $I$ have an impact that is dependent upon the magnitude of $J$.}
\end{figure*}

Similarly, the regularity of a time series can be adjusted by varying the number of slow or fast variables.  As illustrated in Fig.~\ref{changemn} (across columns), increasing $J$ creates a more regular system for all values of $I$.  For $I=6$, the system is fairly regular for both $J=16$ and $J=40$.  However, with only a few fast variables active (e.g. $J=8$), oscillations become increasingly aperiodic.  For $I=8$, additional fast variables are required to achieve the consistent pattern; they allow the system's energy to be more evenly distributed, and nonlinearities in the fast variables have less impact on the stability of the slow variables. \par

Adjusting the number of slow variables greatly alters the qualitative dynamics of the time series for each particular $x_{i}$ (Fig.~\ref{changemn}, across rows).  For $J=8$, only $I=4$ exhibits regularity.  However, with additional fast variables ($J=16$), the $I=6$ system exhibits the greatest regularity.  For $I=4$, the slow variable relative maxima are more pronounced than the relative minima, and the system appears to be quasi-periodic with a dominant frequency of roughly ten days.  For $I=6$, the symmetrical pattern emerges with a frequency of roughly five days.  This pattern is similar when $J$ is increased to $40$, but with decreased amplitude.  However, for $I=4$ and $I=8$ with $J=40$, the pattern being repeated is more complex. \par

We make the preceding observations to motivate the choice of Lorenz '96 as an appropriate testbed with which to perform ensemble forecasting experiments.  The dynamics are quite sensitive to the system parameters, and we will describe forecast results, which vary depending on the underlying dynamics.  \par  

\subsection{Coupling}
\begin{figure}
\includegraphics{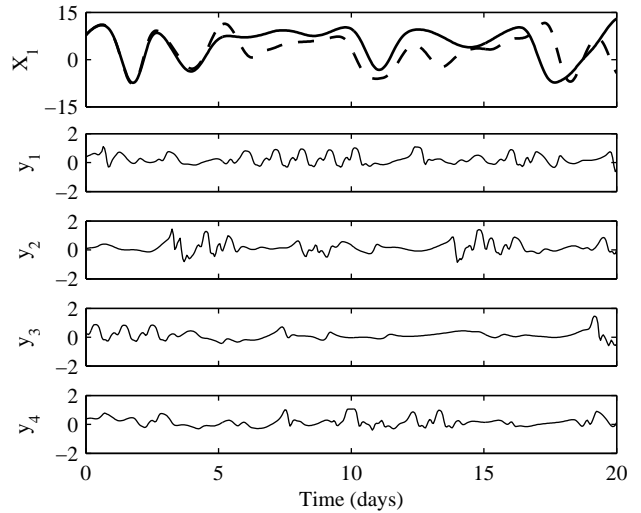}
\caption{\label{series}A $20$ day time series for $x_{1}$ and $y_{1,2,3,4}$ using $I~=~8, J~=~4$.  For the top frame, the solid line represents the time series for $x_{1}$ with fast variable coupling set at $h=1$ (the system).  The dashed line has fast variable coupling reduced to $h=0.5$ (the model).  The bottom four frames are the fast variable time series for the system $y_{1,2,3,4}$ coupled to $x_{1}$, as illustrated in Fig.~\ref{schematic}}
\end{figure}

Each $x_{i}$ in the system is coupled to $J$ small-amplitude fast variables $y_{j}$, and the strength of this coupling is controlled by the parameter $h$.  As mentioned above, the system is defined by setting $h=1$.  However, the model is rendered imperfect by dampening the effect of the fast variables in Eq.~\eqref{fast} by setting $h=0.5$, hereafter referred to as the \textit{model}.  A $20$ day time series for a single $x_{i}$ with its corresponding $y_{j}$'s is shown in Fig.~\ref{series} ($I=8,J=4$).  In the top frame we compare the time series of $x_{1}$ for integrations of the same initial condition by both the system and the model.  By day $20$ the $x_{1}$ values differ by the climatological span of $x_{1}$.  In addition to increasing the amplitude of slow variable oscillations, reducing the coupling $h$ has the effect of creating a less regular time series.  Note that the fast variables (shown for the system) vary with amplitudes on the order of $10\%$ of the slow variables.  \par

\section{Methods}
\begin{figure*}
\subfigure[\label{attr2}Two-dimensional view of orbit segment of system attractor]{\includegraphics{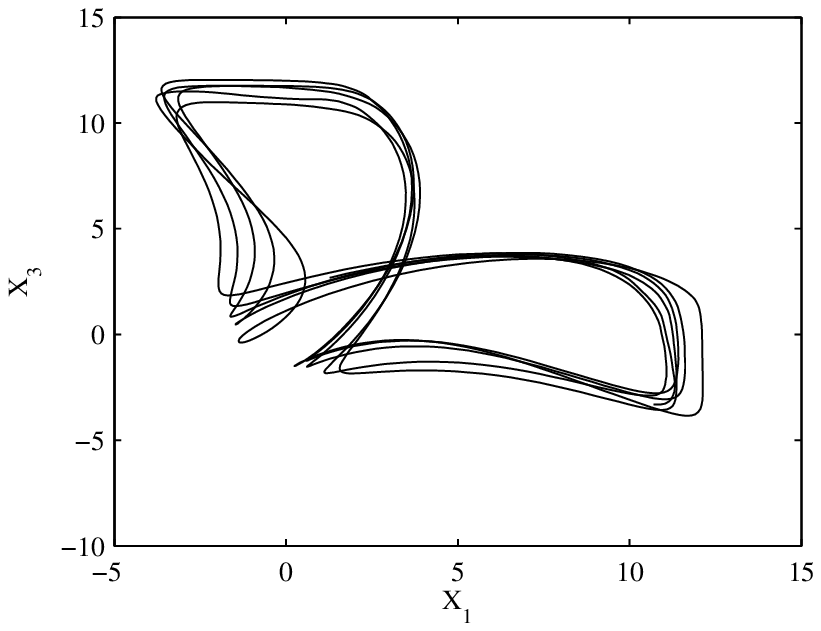}} 
\subfigure[\label{attr3}Three-dimensional view of orbit segment of system attractor]{\includegraphics{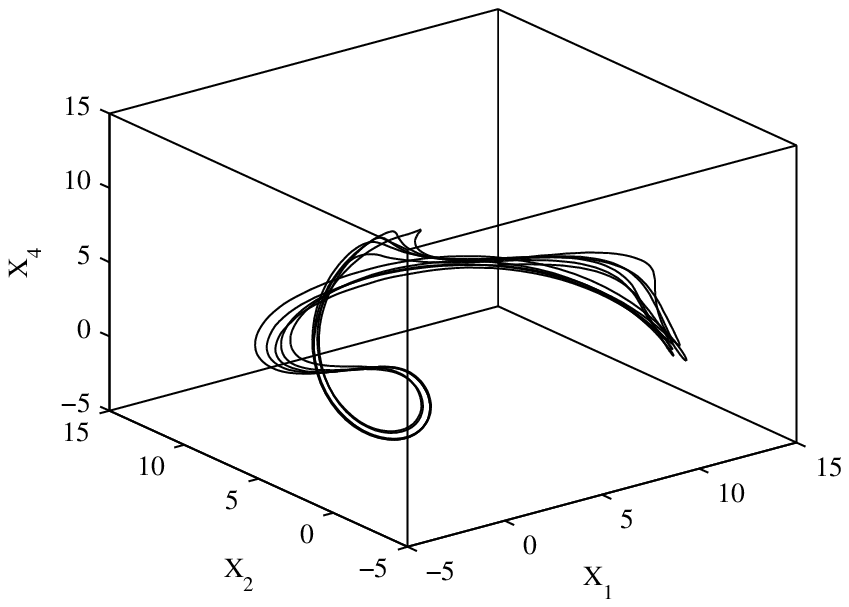}}
\\
\subfigure[\label{forc2}Two-dimensional view of model attractor]{\includegraphics{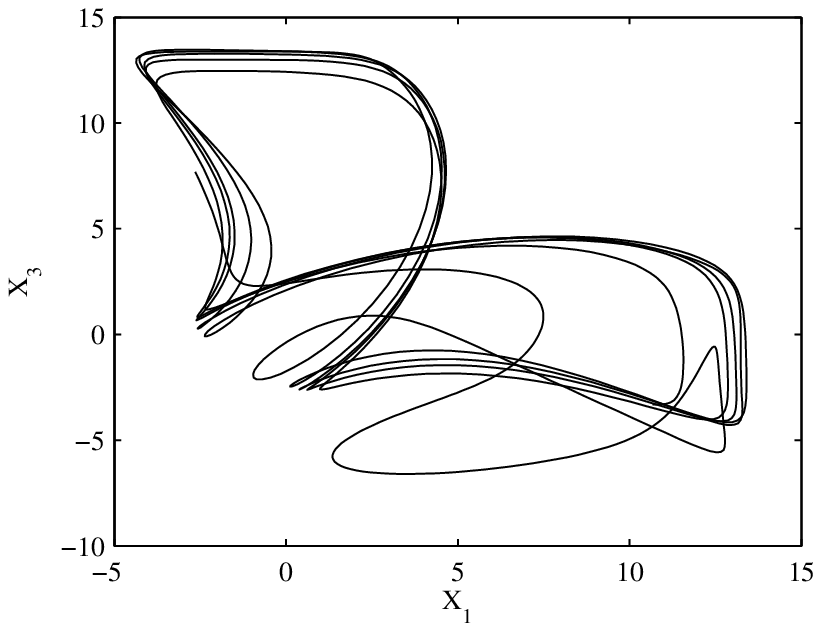}} 
\subfigure[\label{forc3}Three-dimensional view of model attractor]{\includegraphics{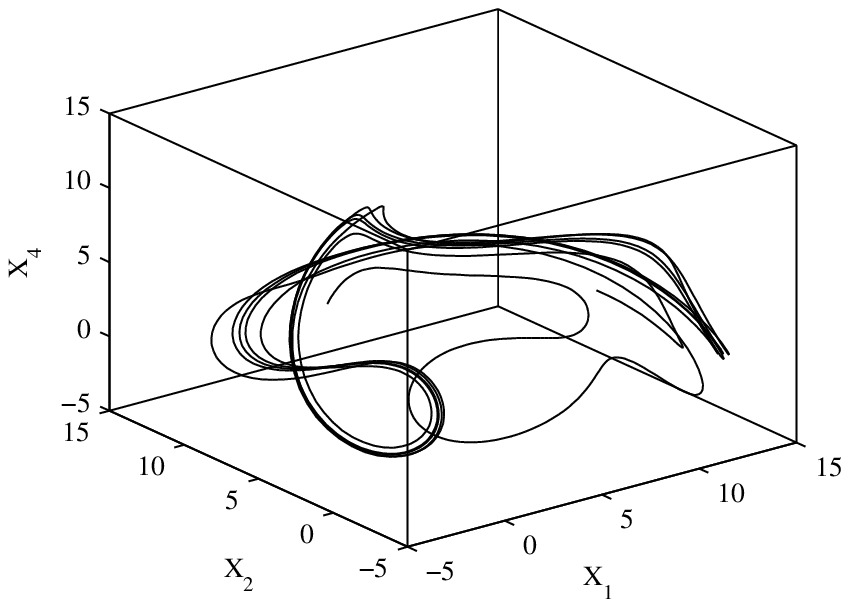}}
\caption{\label{attr}System (a-b) and model (c-d) attractor.  Panels (a) and (c) show two-dimensional views looking at $x_1$ vs $x_3$.  Panels (b) and (d) show three-dimensional views using $x_1, x_2$, and $x_4$ ($I=4,J=16$).}
\end{figure*}

From a given initial condition, the trajectory of the truth ($\textbf{z}^a$) is created by integrating the system in Eqs.\eqref{slow} and \eqref{fast}.  Two- and three-dimensional views of orbit segments of this attractor with $I=4, J=16$ are shown in Fig.~\ref{attr2} and~\ref{attr3}.  As mentioned above, the forecast for each ensemble member is then created by reducing the weight of the fast variables by $50\%$ in the model, relative to the system.  Two- and three-dimensional slices for the model can be seen in Fig.~\ref{forc2} and~\ref{forc3}.  This particular experimental design was chosen as it is typical for global atmospheric models to attempt to parameterize sub-grid scale behavior (e.g. for phenomena occurring on a finer temporal/spatial scale). \par   

\subsection{Ensemble creation} 
The experimental design is inspired by and follows that of Wilks \cite{Wilks05}.  First, long integrations of the system on randomized initial values were performed to establish the shape of the attractor.  A set of $500$ different $I$-dimensional points were then chosen at $250$-day intervals.  This spacing was chosen to ensure that the initial states sampled different regions of the system attractor, and neighboring states were uncorrelated.  A `true' trajectory $\textbf{z}^{a}$ from each of these states was determined using a $50$-day integration of the system.  The goal is to use the model to shadow each of these individual true trajectories with an ensemble of $20$ members.  Care was taken to establish an ensemble that was a dynamically realistic sample of state space near the initial state as follows.\par 

At each of the $500$ initial states, an $I$-dimensional hypersphere was constructed encompassing $100$ neighboring states from the system attractor, identified during a very long system integration.  A neighboring state is defined to be one within $5\%$ of the span of $x_{i}$ in the $i^{\text{th}}$ dimension of the attractor.  For each hypersphere, the covariance of the $100$ neighboring states is calculated, yielding a $I\times I$ matrix $C$.  The matrix values are then scaled to ensure that the average ensemble standard deviation is $5\%$ of the climatological span of the system attractor.  Thus, for each of the $500$ hyperspheres, $20$ initial ensemble members are drawn from the distribution:

\begin{equation}
C^{\text{init}}=\dfrac{0.05^2\tau^2}{\lambda}C,  
\end{equation}
where $\lambda$ is the average eigenvalue of $C$ and $\tau$ is the standard deviation of $x_{i}$ observed in the attractor.  \par 

First, a control state is picked within each hypersphere by adding appropriately distributed random noise to the $I$ slow variables of the truth as follows:
\begin{equation}
\textbf{z}_{0}^{f}(i,1)=\textbf{z}_{0}^{a}(i,1)+\sqrt{C^{\text{init}}}\textbf{y}(i,1),
\end{equation}
for $i=1,2,\dots,I$.  The vector $\textbf{y}$ is $I$-dimensional, consisting of random entries from a Gaussian distribution.  The Cholesky decomposition is used to calculate the square root of $C^{\text{init}}$ \cite{Golub96}.  Note that if $C_{\text{init}}$ was not symmetric positive definite (due to the finite sample in time), then the lower triangular matrix was matched to the upper triangular matrix.  The remaining $19$ ensemble members are then chosen using the same method, but using the control state $\textbf{z}_{0}^{f}(i,1)$ for $i=1,2,\dots,I$ as the central reference point.  Thus, we have
\begin{equation}
\textbf{z}_{0}^{f}(i,m)=\textbf{z}_{0}^{f}(i,1)+\sqrt{C^{\text{init}}}\textbf{y}(i,1),\label{ensemble}
\end{equation}
for $i=1,2,\dots,I$ and $m=2,3,\dots,20$, where $\textbf{z}_{0}^{f}$ is a matrix with ensemble members as columns.  Finally, the initial fast variables for the forecast are set equal to the initial fast variables of the truth, $\textbf{z}_{0}^{f}(j,m)=\textbf{z}_{0}^{a}(j,1)$ for $j=I+1, I+2,\dots,I(j+1)$ and $m=1,2,\dots,20$.  These fast variables are subsequently ignored when measuring the shadowing time.\par

\subsection{Characterizing the ensemble}
Once the ensemble of initial states has been created for each hypersphere, the trajectory of each ensemble member is forecast using the model.  Let $A$ be the $I\times(m-1)$ matrix of ensemble members with the control state at the origin, where $m$ is the number of ensemble members (including the control forecast) and we are only considering slow variables.  If the matrix $A$ acts on every vector in the $(m-1)$-dimensional unit sphere, the result is a filled-in ellipsoid, provided $m>I$.  The points within the ellipsoid can be thought of as representing all potential ensemble members based upon the known $m$ members (all of which lie within this ellipsoid).  Although the system is chaotic, over a short time span one can approximate the nonlinear evolution of the initial sphere of trajectories with this ellipsoid \cite{Alligood96}.  \par 

\begin{figure}
\includegraphics{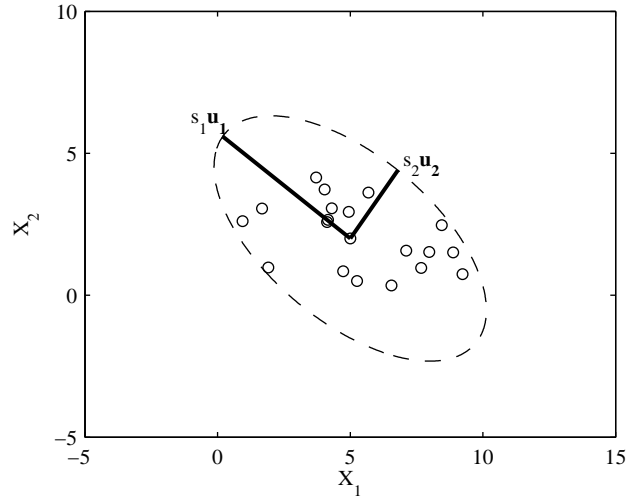}
\caption{\label{ellipsoid}A two-dimensional representation of a $20$-member ensemble $(I=2)$.  The black point is the control forecast, and the white points are the other ensemble members.  Let $A$ be the $2\times19$ matrix of ensemble members with the control forecast defined as the origin.  The semi-axes are calculated using SVD, where $s_{1}$ and $s_{2}$ are the two singular values of the matrix $A$.  Note that the SVD spectrum can be quite steep for forecasts longer than a few time steps.}
\end{figure}

Every two time steps, the ellipsoid is analyzed to determine expanding and contracting directions (i.e. those directions with positive/negative local finite-time Lyapunov exponents).  This frequency was chosen rather than every time step to ensure the identified directions continue to expand/contract.  Let $A=USV^{\top}$ be a singular value decomposition (SVD) of $A$, and let $\textbf{s}(t)~=~s_{1},s_{2},\cdots,s_{I}$ be the singular values (the diagonal entries of $S$) and $U(t)=\left[\textbf{u}_{1},\textbf{u}_{2},\cdots,\textbf{u}_{n}\right]$ the matrix of left-singular vectors at a given time $t$.  Thus the lengths and directions of the semi-axes of the ellipsoid approximately encompassing all potential ensemble members are given by $s_{i}\textbf{u}_{i}$ \cite{Golub96}.   At each time interval, the vector $\textbf{s}(t)$ is compared to $\textbf{s}(t-1)$ (the singular values from the previous time interval) to determine which singular values are decreasing.  We then computed the dot product between all possible combinations of former $u_{i}(t-1)$ and current $u_{i}(t)$ left singular vectors, with the largest dot product chosen to identify direction pairs.  Note that on occasion, a perfect matching was not achieved, but the best guess was used.  A two-dimensional analog is shown in Fig.~\ref{ellipsoid}.  Note all ensemble members lie within the ellipse, and the control state is located at the center by definition of A.  SVD allows us to identify the ellipse semi-axes $s_{1}\textbf{u}_{1}$ and $s_{2}\textbf{u}_{2}$.  \par 

\begin{figure}
\includegraphics{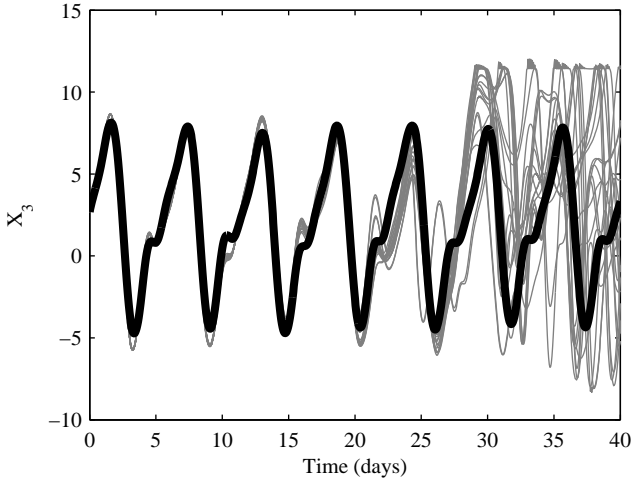}
\caption{\label{typical}A $20$-member ensemble forecast for $x_{3}$ initialized using Eq.~\eqref{ensemble}.  The black line represents the truth, an integration of the system.  The gray lines are individual ensemble members, integrations of the model ($I~=~6, J~=~16$).}
\end{figure}

An example one-dimensional $40$ day ensemble forecast is shown in Fig.~\ref{typical} ($I=6, J=16$).  Note that for the first $20$ days, almost all ensemble members remain close to the truth.  However, by day $40$ the spread of ensemble members covers the entire state space, and the ensemble mean no longer accurately represents the truth.  In addition, Fig.~\ref{typical} illustrates that the model achieves greater extreme values than the system.  With less energy being dissipated through the fast variables, more energy remains oscillating between the slow variables in the model, as compared to the system.  Note that the reduced regularity of the model relative to the system can also be observed.  \par

\section{Forecasting}
\begin{figure}
\includegraphics{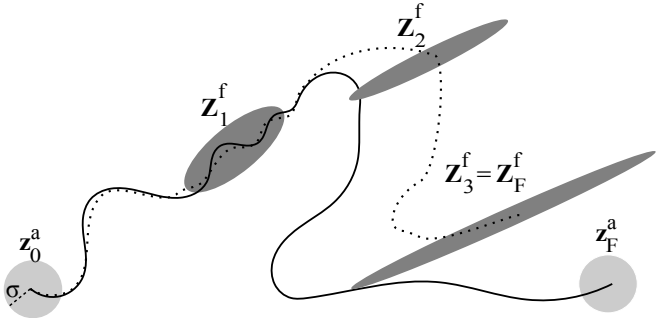}
\caption{\label{forccart}Schematic of forecasting in two dimensions.  The black line represents the trajectory of the truth ($\textbf{z}^{a}$).  $\textbf{Z}^{f}_{t}$ is the ellipsoid encompassing the ensemble members at time $t$, and the dashed line is the trajectory of the ensemble mean.  For nonlinear systems, this ellipsoid is only a good approximation of the collections of forecasts for small $t$.  Note that the time steps $0, 1, 2, 3$ are an incomplete sampling of all time steps.  The goal of a forecast is to achieve an ensemble which has a nonempty intersection with $\textbf{z}^{a}_{F}$.  Here forecasting fails after time $2$.}
\end{figure}

Forecasting often fails when the state space dynamics of the physical system expand along a previously contracting dimension.  A two-dimensional illustration of this concept is shown in Fig.~\ref{forccart}.  Because of the uncertainty associated with the initial location of the truth, an $I$-dimensional sphere of radius $\sigma$ is used to represent the truth.  It is from within this sphere that the initial ensemble members are chosen.  The ellipse representing the ensemble members diverges from the trajectory of the truth when the system quickly turns in a new direction.  In the illustration, forecasting has failed at time $F$ as there is no overlap between the $I$-dimensional sphere around the truth $\textbf{z}^{a}_{F}$ and the ensemble ellipsoid $\textbf{Z}^{f}_{F}$.  For all experiments, the number of expanding directions observed throughout the forecast was non-constant.  This unstable dimension variability is a cause of shadowing failures well-documented in the literature \cite{Danforth06, Dawson94, Yuan00, Sauer02}. \par

\subsection{Forecasting with inflation}
\begin{figure}
\includegraphics{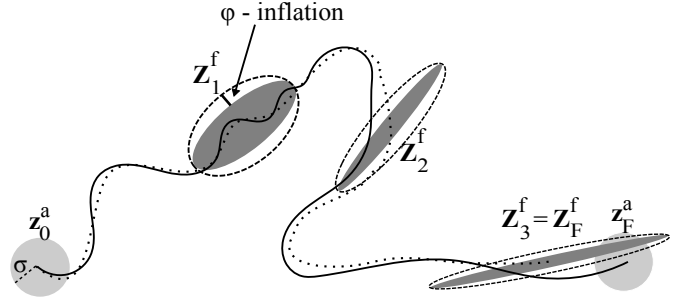}
\caption{\label{infcart}Schematic of forecasting with inflation in two dimensions.  The black line represents the trajectory of the truth.  $\textbf{Z}^{f}_{t}$ is the ellipsoid encompassing the ensemble members at time $t$, and the dashed line is the trajectory of the ensemble mean.  At each time step, the ellipsoid is inflated by $\varphi$ along contracting directions, where in this case there is only one such direction.  Note that forecasting now achieves its goal in contrast to Fig.~\ref{forccart}.}
\end{figure}

As described in the introduction, inflation is performed in an attempt to improve the ensemble forecast by expanding the ellipsoid along contracting dimensions of the state space.  For each direction $\textbf{u}_{i}$ in which the ensemble is contracting, ensemble members are inflated along the $i^{\text{th}}$ semi-minor axis of the ellipsoid $s_{i}\textbf{u}_{i}$.  Although this adds artificial uncertainty to the forecast, Danforth and Yorke argue that the amount is minimal \cite{Danforth06}.  If the ellipsoid continues to contract in these dimensions, the synthetic uncertainty continuously decreases.  However, if the system begins to expand along a previously contracting dimension, the inflated ensemble will capture some of the change.  This concept is illustrated for two dimensions in Fig.~\ref{infcart}.\par 

As described above, the directions of the semi-axes are calculated using SVD throughout the forecast at periodic time intervals and $\textbf{s}(t)$ is compared to $\textbf{s}(t-1)$.  For each $i$ in which $s_{i}<s_{i}(t-1)$, all ensemble members are inflated in direction $\textbf{u}_{i}$.  It is important to note here that if inflation were performed in expanding dimensions as well, forecasts would degrade more rapidly.  By constraining inflation to the contracting dimensions, we are minimizing the damage caused in hyperbolic regions of state space.  \par 

Let $\varphi$ be the inflation amount, expressed as a percentage of the semi-axis length, and let $A$ be the matrix of ensemble members with the control state at the origin, prior to inflation.  Define \par
\begin{equation}
M=\mathbb{I}+\varphi(\textbf{u}_{i} \textbf{u}_{i}^{\top})
\end{equation}
where $\mathbb{I}$ is the $I$-dimensional identity matrix, $\textbf{u}_{i}^{\top}$ is the transpose of $\textbf{u}_{i}$, and let $A'=MA$.  Adding the control state to each vector in $A'$ then yields ensemble members inflated by $\varphi$ in direction $\textbf{u}_{i}$.  \par 

\begin{figure}
\subfigure[No inflation]{\includegraphics{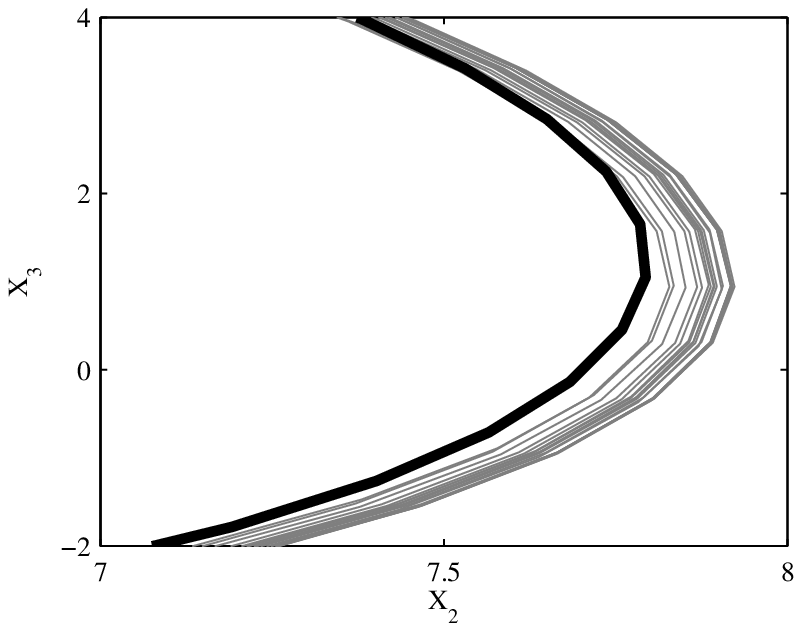}}
\\
\subfigure[Using inflation]{\includegraphics{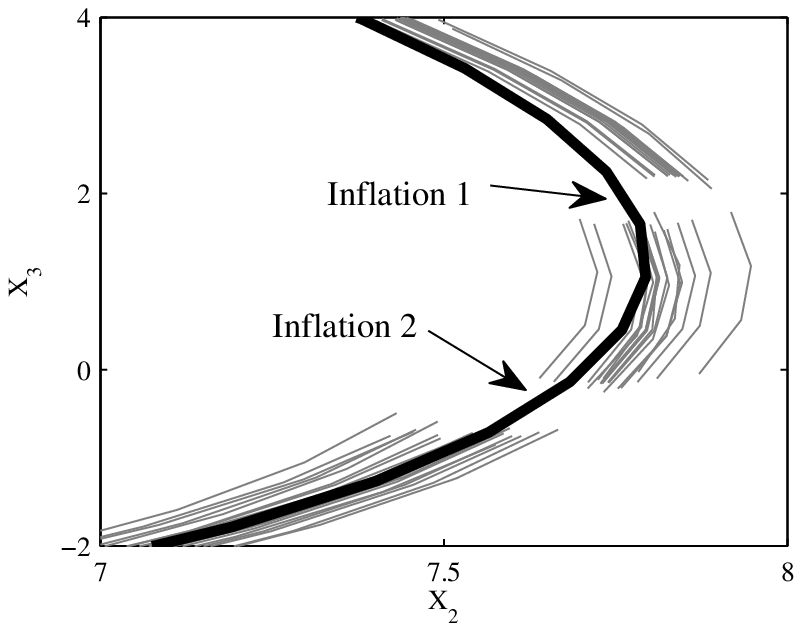}}
\caption{\label{Infworks}Successful inflation in an ensemble forecast.  The black line is the true trajectory $\textbf{z}^{a}$, and the gray lines are ensemble member forecasts without inflation (panel a) and with inflation having the potential to occur at a minimum of every two time steps (panel b).  All trajectories are moving in a clockwise direction.  At location $1$, forecasting fails without inflation.  Location $2$ represents a second inflation of the ensemble.  Inflation successfully modifies the ensemble to include the truth as a plausible member.  Note that no observations of $\textbf{z}^{a}$ are used during the inflation at locations $1$ and $2$ ($I=6, J=16$).}
\end{figure}

The benefit of inflation for a particular forecast is illustrated in Fig.~\ref{Infworks} for a two-dimensional projection of an ensemble forecast ($I=6$).  As the trajectory of $\textbf{z}^{a}$ begins to turn clockwise, the forecasts (gray lines) diverge from the black line.  As represented by the gray lines, forecasting fails at location $1$ without inflation (Fig.~\ref{Infworks}a), when there is no longer overlap between the $\sigma$-sphere surrounding $\textbf{z}^{a}$ and the ensemble ellipsoid.  However, in Fig.~\ref{Infworks}b, the gray lines are inflated at location $1$ in previously contracting directions (as inferred from SVD of the model ensemble).  Note that the gaps in the figure represent both an inflation and a time step.  As the figure illustrates, these shifts help capture the change in direction for $\textbf{z}^{a}$.  It is important to note that no measurements of $\textbf{z}^{a}$ are used to determine the directions in which to inflate.  A second inflation for the gray lines occurs four time steps later at location $2$ in panel b.  At first there is more uncertainty in the ellipsoid encompassing the ensemble members.  However, as $\textbf{z}^{a}$ continues in the same direction, the uncertainty is dampened by the contracting dynamics.  Note that although the ellipsoid is analyzed every two time steps, the model did not inflate between locations $1$ and $2$.  Also, inflation acts symmetrically on an ensemble, leaving the ensemble mean unchanged during a single step.  However, once the ensemble is integrated forward in time, the inflated and non-inflated ensemble mean forecasts diverge.  \par   

\par

\subsection{Results}
For each of the $500$ initial hyperspheres, a $50$-day model forecast for the $20$ ensemble members was created with and without inflation.  At each point in time, let $\textbf{z}^{*}$ be the mean of the $20$ ensemble members' slow variables.  For each hypersphere, $\textbf{z}^{*}$ was compared to the slow variables of the truth, where we let $\textbf{z}^{a}$ represent only the slow variables for simplicity.  The root mean square error (RMSE) and anomaly correlation (AC) were calculated at each time step as follows:
\begin{equation}\label{RMSE}
RMSE=\lVert\textbf{z}^{*}-\textbf{z}^{a}\rVert_2
\end{equation}
\begin{equation}\label{AC}
AC=\dfrac{(\textbf{z}^{*}-\overline{\textbf{z}})\cdot(\textbf{z}^{a}-\overline{\textbf{z}})}{\lVert \textbf{z}^{*}-\overline{\textbf{z}}\rVert_2 \lVert \textbf{z}^a-\overline{\textbf{z}}\rVert_2}
\end{equation}
where $\overline{\textbf{z}}$ represents the vector of system climatological averages for each slow variable.  RMSE and AC calculations were then averaged over all hyperspheres for the duration of each forecast.  For the system, RMSE $\approx 9$ represents saturation, at which point there is no similarity between the forecast and the truth.  \par 

\begin{figure}
\includegraphics{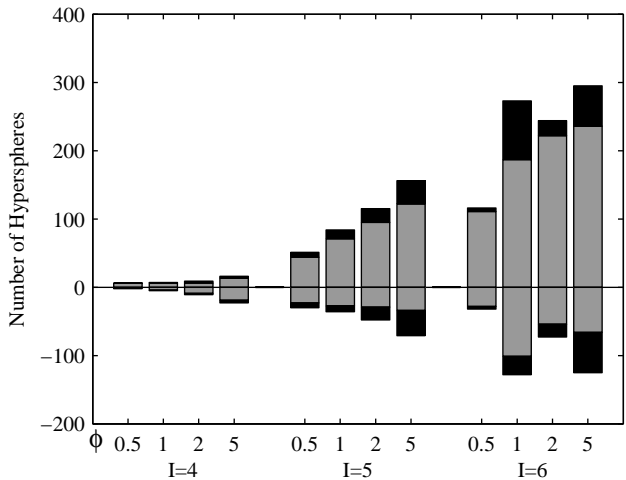}
\caption{\label{forecasts}Inflation forecasting results for $500$ hyperspheres.  Bars above zero show the number of hyperspheres for which inflation improved the forecast, while bars below zero show the number of hyperspheres for which inflation degraded the forecasts.  Black bars indicate improvements and degradations of the average shadowing time by more than $5\%$ (``inflation succeeded" and ``inflation failed," respectively).  Gray bars represent the categories ``inflation helped" and ``inflation hurt."  The number of times inflation improved the forecast generally exceeded the number of degradations, both of which tend to increase with dimensionality and $\varphi$.}
\end{figure}

For the present discussion, we consider the period for which a forecast is useful, namely the time for which AC remains above $0.6$ \cite{Kalnay03}.  For each state space size, the average useful time was calculated using the non-inflated forecasts.  This yielded durations of $6.7, 8.7,$ and $17.0$ days, corresponding to $I=4, 5,$ and $6$, respectively.  If the inflated and non-inflated forecasts crossed the $0.6$ threshold during the same time step, the useful times were considered indistinguishable.  Inflation was deemed to have ``succeeded" if the duration for an acceptable forecast improved by more than $5\%$ of the average useful time for $\varphi=0$ forecasts.  Similarly, the technique was considered to have ``failed" if the duration worsened by the same threshold.  If there was any improvement (greater than one time step) with inflation, forecasts were denoted ``inflation helped."  Any measurable degradation with inflation was labeled ``inflation hurt."  Results for $I=4, 5,$ and $6$, with $\varphi=0.5\%, 1\%, 2\%,$ and $5\%$, are recorded in Fig.~\ref{forecasts} with totals out of $500$.  Bars above zero represent improvements, while bars below zero represent degradations.  Black bars indicate improvements and degradations of more than $5\%$ of the average useful time (i.e. the number of successes and failures).  Other metrics such as RMSE and the trajectory of the best ensemble member were considered, but they showed the same general trends.  For all experiments described by Fig.~\ref{forecasts}, $J=16$  as illustrated in Fig.~\ref{changemn}.\par

\par

Although the sample size is small, Fig.~\ref{forecasts} suggests that inflation has a greater effect for larger $I$, with the best results exhibited for $I=6$ and $\varphi=1\%$.  In fact, the number of hyperspheres in which inflation helped was an order of magnitude better for $I=5$ than for $I=4$.  The vast majority of inflated $I=4$ forecasts were identical to non-inflated ($\varphi=0$) forecasts.  These numbers roughly doubled for $I=6$ relative to $I=5$, likely due to the increased regularity in $\textbf{z}^a$  for $I=6$ observed in Fig.~\ref{changemn}.  Unfortunately, the same trend is seen for forecasts in which inflation failed.  By capturing one of the sudden changes in the trajectory of the truth using inflation, the model can better forecast the truth for a longer duration.  Alternatively, this increase in the occurrence of inflation helping might be evidence that error introduced by inflation in contracting directions has a minimal effect, as argued by Danforth and Yorke \cite{Danforth06}.  This effect would likely be strongest in systems with a steep singular value spectrum, where the expanding dimensions play a more dominant role. \par 

Naturally, the success of inflation is also strongly dependent on $\varphi$, with increasing $\varphi$ generally corresponding to a larger number of both successes and failures.  By increasing the inflation amount, the ellipsoid encompassing the ensemble members is more likely to overlap with the trajectory of the truth, and thus capture unstable dimension variability events.  On the other hand, inflation can make a forecast worse.  By inflating in directions incommensurate with the truth, the method introduces additional error, which can be quite significant when amplified by nonlinearity.  The number of hyperspheres in which inflation failed also naturally increases with $\varphi$.  However, as mentioned above, provided the contracting directions continue to contract, the error introduced is hopefully minimal. \par 

For a given set of experimental parameters, the averaged useful time over all $500$ hyperspheres is used for analysis.  However, the $\varphi=0$ forecast provides an individual useful time for each hypersphere.  Whether or not inflation helps or hurts (i.e. reduces or increases the individual useful time) is not dependent upon the magnitude of the individual useful time.  For example, inflation is equally likely to help (or hurt) an individual forecast whose non-inflated useful time was $3$ days or $20$ days. \par 

\begin{figure}
\subfigure[RMSE]{\includegraphics{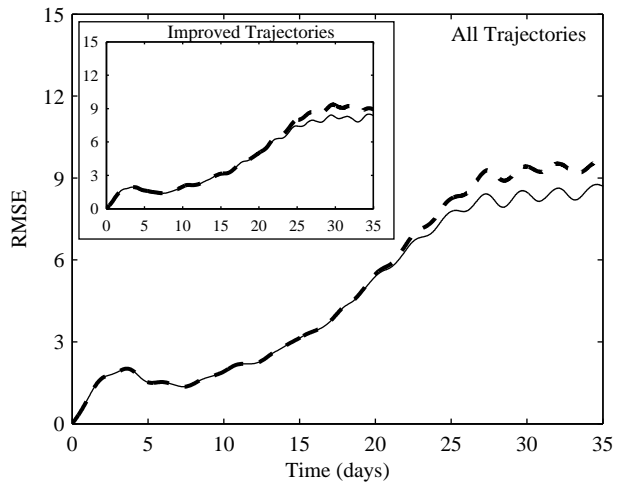}}
\\
\subfigure[AC]{\includegraphics{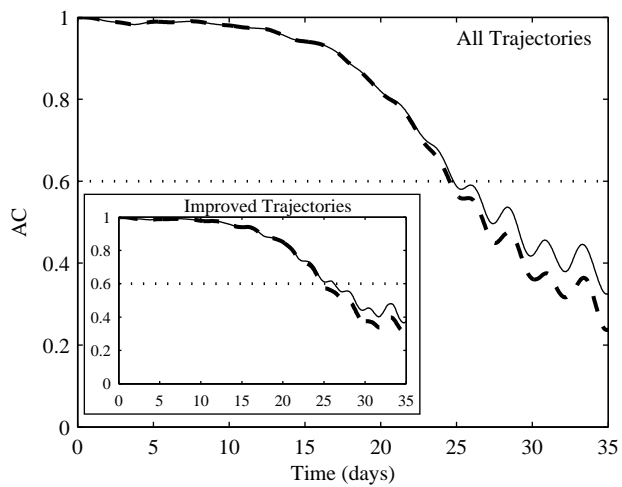}}
\caption{\label{Goodones}Averaged RMSE (a) and AC (b) for the $I=6, J=16$ system.  The dashed line represents no inflation, while the solid line is with inflation of $1\%$.  Figures show the averaged data for all $500$ hyperspheres, while insets only include the hyperspheres for which inflation succeeded.  Given the small (positive) change in average forecast statistics, the method is unlikely to harm forecasts, and may be adjusted to isolate particular events for which inflation can be quite successful (see Section VI).}
\end{figure}

As shown in Fig.~\ref{forecasts}, for many of the hyperspheres inflation successfully improved the duration for which a forecast is useful.  The lack of harm caused by this technique in a global sense can be seen by averaging RMSE and AC over the hyperspheres in which inflation was successful as shown in Fig.~\ref{Goodones} for $I=6, J=16$.  Both plots show small improvement relative to forecasts made with no inflation.  In section VI, we demonstrate the ability to isolate conditions likely to result in successful inflation as in Fig.~\ref{Infworks}.\par

Conducting experiments for larger dimensional systems was difficult due to computation time limitations for estimating the local dynamics of the system attractor (see Section III).  To reduce the computational burden, initial ensemble members were chosen using a spherically-symmetric Monte Carlo approach (i.e. nearby initial conditions but with no attractor-imposed structure) for these larger systems ($I=7, 8, \dots, 40$).  The $I=6, I=16,$ and $I=21$ systems had substantially greater average useful times than the other systems (by a factor of five).  These specific systems all exhibit increased regularity in their time series (see Fig.~\ref{changemn}), presumably a result of some resonance between the slow and fast variables.  For these systems the model is most sensitive to inflation, with small $\varphi$ greatly improving forecasts and larger $\varphi$ greatly degrading forecasts.  The less-regular systems had smaller incidence of success and failure, with results similar to those observed for the $I=4$ and $I=5$ systems.  However, the use of a spherically-symmetric Monte Carlo approach to initialize an ensemble degraded the performance of inflation when compared to the more sophisticated ensemble initialization technique.  \par 

\section{Shadowing and Stalking}
\subsection{Methods}
\begin{figure}
\includegraphics{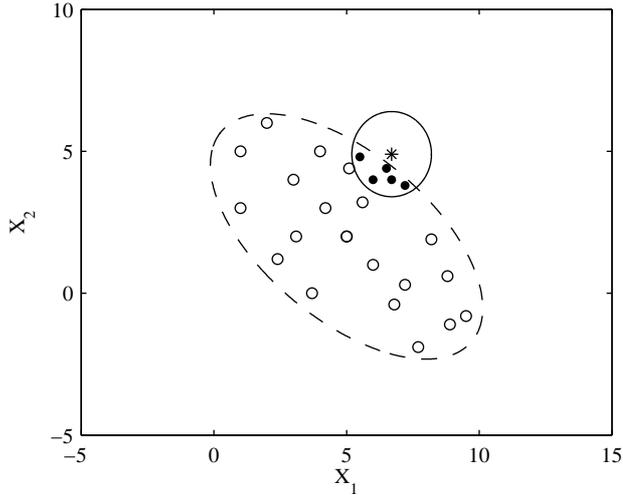}
\caption{\label{redefining}Schematic of the redefining of a shadowing ensemble in two dimensions.  The star represents the known location of the truth.  The solid circles are supplemented by a collection of points meant to approximate the overlap of the ellipse and the sphere around the truth.  The open circles are previous ensemble members that will now be redefined within the overlap, which is approximated by a new ellipsoid.  Note that the $x_{1}, x_{2}$ axes are exaggerated for visual ease.}
\end{figure}

Shadowing with an ensemble presents an alternative method for assessing the quality of a model's predictions.  Contrasting the forecasting experiments described previously, in this context the trajectory of the truth is known a priori to within a given uncertainty ($\sigma$), and its location at a given time is represented by an $I$-dimensional sphere of radius $\sigma$.  At regular intervals, the intersection between the ellipsoid encompassing the ensemble members and the $\sigma$-sphere is approximated.  The ensemble ellipsoid is then redefined to represent the intersection, while the other ensemble members are redefined to fall within the overlap.  Provided this intersection is nonempty, the trajectories of some ensemble members yield accurate representations of reality.  A schematic of this technique is shown in Fig.~\ref{redefining}.  Note that although the actual truth lies outside the ensemble ellipsoid, some ensemble members are within $\sigma$.  The trajectories of these ensemble members can be forecast forward in time, and the process is repeated.  If some ensemble members remain in the intersection for the entire forecast, then their trajectories have successfully $\sigma$-shadowed the truth.  When inflation is applied to the ensemble ellipsoid (as described previously), this aggressive form of shadowing is called \textit{stalking}.  \par 

\subsection{Results}
\begin{figure}
\includegraphics{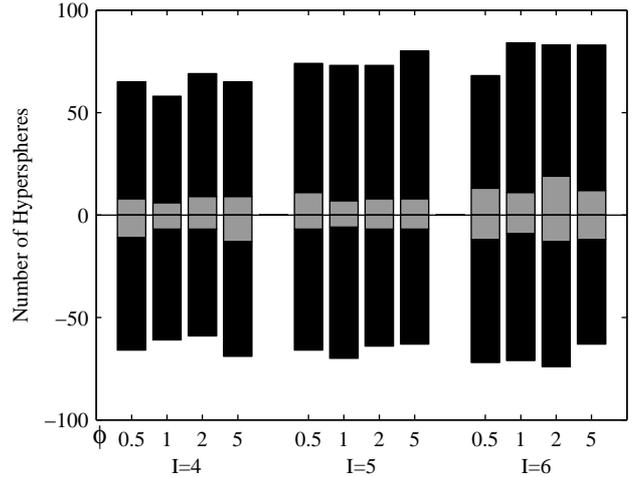}
\caption{\label{stalking}Stalking results for $500$ hyperspheres.  Bar categories are identical to those in Fig.~\ref{forecasts}.  Inflation improves and worsens the ability to shadow with similar probability.  Counts tend to slightly increase with the dimension of the system.}
\end{figure}

The same initial conditions utilized in the forecasting experiments were used to create $50$-day stalking ensembles.  However, every four time steps the ensemble ellipsoid was redefined as described in Fig.~\ref{redefining}.  This frequency was chosen to be long enough to ensure that appropriate contracting dimensions could be identified.  During the definition of the overlap between the ensemble ellipsoid and the truth sphere, semi-axes can be altered drastically, requiring several time steps to reflect the local dynamics.  The $\sigma$-shadowing distance (uncertainty around the truth) was chosen to be $10\%$ of the climatological range in each dimension, namely twice the spread in the initial ensemble.  A new collection of $20$ ensemble members were then chosen lying within the redefined ellipsoid.  As before, RMSE and AC were calculated and compared for both shadowing and stalking experiments.  Counts for the number of hyperspheres for which inflation succeeded, failed, helped, and hurt as described in Section IV were tallied.  Results for $I=4, 5,$ and $6$, with $\varphi=0.5\%, 1\%, 2\%,$ and $5\%$, are given in Fig.~\ref{stalking}.  \par 

\par

The shadowing experiments exhibited a greater benefit from inflation when the dimensionality $I$ is large, although the difference is less than before.  However, the number of hyperspheres for which inflation succeeded does not seem to trend monotonically with $\varphi$ over the range $0.5\% - 5\%$.  Since the number of hyperspheres in which inflation succeeded roughly equals the number for which inflation helped, we can conclude that when stalking is constructive, it improves the shadowing time by more than one day.  On the other hand, the number of hyperspheres in which inflation succeeded is nearly equivalent to the number for which inflation failed for all experiments.  Thus, inflation can also have a negative effect upon shadowing.  \par 

As with forecasting, stalking experiments were extended to higher dimensional systems using a spherically-symmetric Monte Carlo approach for ensemble initialization.  Similar to the lower dimensional systems in Fig.~\ref{stalking}, counts for the number of hyperspheres for which inflation succeeded and failed were relatively close.  For small values of inflation, there were often errors in the procedure by which corresponding ensemble semi-axes directions are calculated between time steps, resulting in an increase in the number of failures.  With larger inflation, the algorithm can better determine corresponding directions during redefinition of the ensemble.  The highly regular $I=6, I=16,$ and $I=21$ systems were less distinguished for stalking experiments.  Since  the truth is known throughout, ensemble members are forced by the redefinition to remain nearby regardless of the regularity of the attractor. \par  

\section{Targeted Inflation}
Despite taking advantage of local stability estimates, inflation as defined previously was naive in the sense that it occurred in all contracting directions, including those which push forecasts away from the truth.  Thus, this naive inflation had the potential to both improve and degrade a forecast.  The final objective of this research was to isolate the instances in which inflation is beneficial to the forecast, and identify the systematic tendencies of these situations.  With this knowledge, at each time step the decision of whether inflation should occur, and in which contracting directions, could be based upon the characteristics of the local state space of the system.  For these \textit{targeted inflation} experiments, the $I=5$ system was chosen for further study; it does not have the same regularity as the $I=6$ system (Fig.~\ref{changemn}), but is more regular than some of the higher dimensional systems. \par 

For a given time step in the forecast, the shape of the attractor near the current state $\textbf{z}^{f}_{t}$ was approximated as follows.  First, the state $50$ time steps prior in the forecast $\textbf{z}^f_{t-50}$ was considered.  By adding uniform random noise in every direction on the order of $\sigma$, $1000$ analogs located near (within $\sigma$) $\textbf{z}^{f}_{t-50}$ were chosen.  These $1000$ analogs were forecast forward $50$ time steps, and the covariance matrix associated with $\textbf{z}^{f}_{t}$ was calculated.  The eigenvectors of this forecast state covariance matrix yield an approximation of the shape of the attractor near $\textbf{z}^{f}_{t}$.  Finally, the dot product of these eigenvectors and the proposed inflation direction (determined by SVD to be contracting) was calculated.  The larger this projection, the more likely the proposed inflation direction is consistent with the local shape of the attractor.  If the resulting dot product was above a set threshold $\mu$ and the direction was determined to be contracting, then inflation proceeded as before.  Otherwise, no inflation occurred.  For all values of $\varphi$, $\mu$ was set at $0, 0.8,$ and $0.9$, reducing the number of inflations by roughly $0\%, 60\%,$ and $80\%$, respectively. \par 

The goal here is to avoid inflating in contracting directions likely to harm the forecast.  By initializing the analogs randomly, we are able to sample a new collection of state space directions, distinct from those spanned by the inflated ensemble. \par

\subsection{Forecasting}
\begin{figure}
\includegraphics{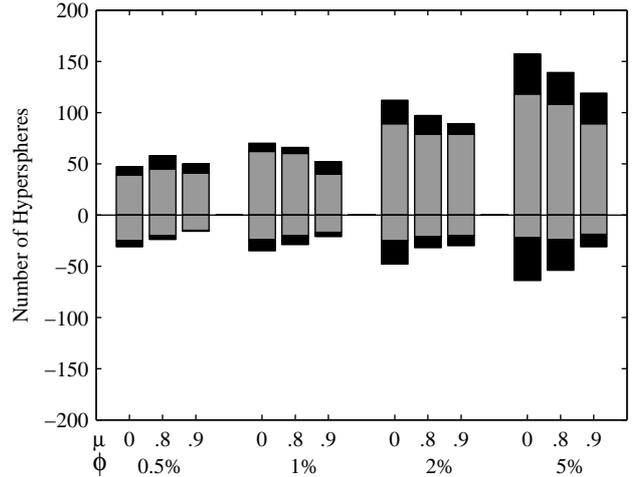}
\caption{\label{decide}Targeted inflation for forecasting experiments on $500$ hyperspheres.  Bar categories are the same as those defined in Fig.~\ref{forecasts}.  Note that $\mu=0.9$ substantially reduces the instances of failure, while not significantly reducing the instances of success ($I=5$).}
\end{figure}

Forecasting results are given in Fig.~\ref{decide}.  Adjusting the number of inflations throughout the forecast by changing $\mu$ clearly improves the method.  Note first that results from a cutoff of $\mu=0$ agree with Fig.~\ref{forecasts}.  The small differences (most notable for $\varphi=1\%$) can be attributed to starting from different initial points in the attractor.  For $\varphi=2\%$ and $\varphi=5\%$, increasing $\mu$ has the effect of decreasing the counts; with fewer inflations, ensembles more closely resemble the non-inflated forecast ($\varphi=0$).  It is important to note that although the number of hyperspheres for which inflation succeeds remains relatively constant, the number for which inflation fails is substantially reduced as $\mu$ is increased.  This trend indicates that the technique is successfully determining when inflation will indeed be useful, and is inflating less frequently when it is likely to worsen a forecast.  For example, consider $\varphi=5\%$ in Fig.~\ref{decide}.  With naive inflation ($\mu=0$), the technique succeeded for $39$ hyperspheres and failed $42$ times.  Setting $\mu=0.9$ reduced the number of successes by $23\%$ to $30$, however, it eliminated $71\%$ of the failures, leaving only $12$.  When $\mu$ is increased beyond $0.9$ though, inflation occurs too rarely and the number of hyperspheres for which inflation succeeds is greatly reduced (not shown).  \par 

\par

For $\varphi=0.5\%$ and $\varphi=1\%$, the trend is a little more complicated.  Still, $\mu=0.9$ resulted in a remarkably better ratio of number of hyperspheres succeeding versus failing.  With $\varphi=0.5\%$, $\mu=0.8$ surprisingly increased the number of hyperspheres succeeding and helping relative to $\mu=0$.  For $\varphi=1\%$, $\mu=0.8$ had notably few hyperspheres for which inflation succeeded.  This indicates that although experiments were run for $500$ different hyperspheres, there is still variability between trials based upon initial conditions and chaotic system dynamics.  

\subsection{Stalking}
\begin{figure}
\includegraphics{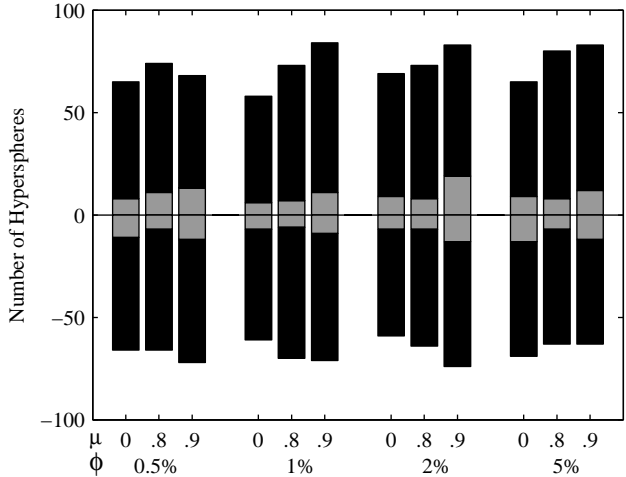}
\caption{\label{decidestalk}Targeted inflation for stalking experiments on $500$ hyperspheres.  Bar categories are the same as those defined in Fig.~\ref{forecasts}.  Results varied, with more work needed to optimize the success of inflation for stalking experiments ($I=5$). }
\end{figure}

As shown in Fig.~\ref{decidestalk}, the results for stalking experiments are more variable.  For some algorithm parameters, such as $\mu=0.8$ for $\varphi=0.5\%$ and $\varphi=1\%$, targeted inflation appears to successfully increase the number of hyperspheres for which inflation succeeded, while decreasing the number for which inflation failed.  However, the same $\mu$ for greater inflation had smaller differences, even increasing the number of trajectories for which inflation failed.  One possible explanation for the mixed results in stalking experiments is the experimental design.  Redefining the ensemble every fourth time step reduces the ability to identify expanding/contracting dimensions.  This can lead to difficulties in matching ellipsoid semi-axis directions from one time step to the next.  Adjusting $\mu$ did alter the results, however more work is needed to optimize the success of inflation for stalking experiments.  \par 

\par

\section{Discussion}
Using the Lorenz '96 coupled system as an analog for atmospheric dynamics, we examined the potential for using inflation to improve model forecasts of the Earth's atmosphere.  In the first experiments, forecasts were made with and without inflation, and the resulting trajectories were compared to the system truth.  In the second part of this work, the existence of a shadowing trajectory with and without inflation was assessed.  The trajectory of the truth was known a priori, and only the closest ensemble members were considered every few time steps.  In the former, the technique is referred to as inflation, and in the latter, stalking.  Under idealized conditions, inflation was shown to be beneficial for both techniques.  The final part of the work attempted to isolate the ensemble directions most likely to benefit from increased uncertainty using targeted inflation. \par           

For the Lorenz '96 model, by adjusting the number of slow and fast variables, and the degree of coupling between them, one could vary between regular and highly chaotic systems; the greatest regularity was observed for $I=6$ and $I=16$.  However, by increasing the number of fast variables ($J$), greater regularity can be achieved for a fixed number of slow variables.  In contrast to hyperbolic systems, the present system exhibits unstable dimension variability and fluctuating Lyapunov exponents as is common in higher dimensional systems \cite{Sauer02}.  \par 

For forecasting experiments, inflation showed the potential to increase the length of time for which an ensemble forecast represents the verifying truth as a plausible member, as illustrated in Fig.~\ref{Infworks}.  Inflation in directions away from the truth, however, can degrade a forecast.  As a result, greater inflation amounts always correspond to an increase in the number of hyperspheres for which inflation failed.  For regular systems such as $I=6$ and $I=16$, smaller inflation amounts can often successfully improve a forecast, while the error introduced from larger inflation amounts is magnified.  However, for less regular systems, larger inflation was necessary to improve forecasts.  For these systems, the uncertainty introduced was less likely to have a harmful effect.  \par 

Stalking experiments were used to establish the existence of a model trajectory that remains close to the truth for a given time, and assess how inflation affects this trajectory.  For these experiments, results were relatively constant for systems with greater than six dimensions.  Inflation was shown to be successful at about the same rate at which it hindered shadowing.  Thus, the utility of stalking is dependent upon the ability to isolate the hyperspheres for which inflation helped. \par 

Finally, by quantifying the local dynamics at each time step and restricting inflation to directions consistent with the local attractor, the instances for which inflation helped were isolated with moderate success for forecasting experiments.  The number of hyperspheres for which inflation failed could be decreased without significantly decreasing the number for which inflation succeeded.  Stalking experiments with targeted inflation yielded mixed results, with additional modifications needed to attain the same degree of success.  In practice however, the truth is not known a priori.  Thus, the improvements observed with targeted inflation for forecasting experiments are encouraging for application in NWP.  \par 

When shadowing physical systems such as the Earth's atmosphere, one is presented with the challenges of uncertainty in the initial condition, sensitive dependence, and model error.  Model error is estimated to dominate the forecast error in weather systems for the first three days \cite{Orrell01}.  Nevertheless, modelers must attempt to provide the best predictions possible given the circumstances.  This research demonstrates that inflation has the potential to aid in our attempts to model chaotic physical systems, provided we are able to isolate the local shape of the attractor and tune the inflation parameters successfully.  Targeted inflation is most likely to be beneficial for highly nonlinear events (e.g. hurricanes), where sudden changes in the state space trajectory are prevalent.  Further exploration of the design parameters of targeted inflation are likely to result in increased effectiveness.\par 

One potential application of this technique to NWP is in maintaining the spread of a forecast ensemble.  For unbiased ensembles, an effective spread is equal to the RMSE of the ensemble mean with respect to the analysis \cite{Wilks11},  but NWP ensembles typically fail to retain this ideal beyond the first few days.  Of the various means by which one could increase the ensemble spread artificially, targeted inflation preserves the ensemble orientation relative to the local model attractor and has the potential to improve the ensemble mean.  \par 

The major disadvantage of targeted inflation, relative to the naive version, is the frequently required computation of the shape of the attractor.  For NWP models, where the number of degrees of freedom is $O(10^{10})$, this computation would be prohibitive.  However, the operational forecasting centers compute new analyses and ensemble forecasts roughly every six hours.  Thus, the required information for an estimate of the attractor shape would be available cost free in terms of computational time. \par 

\section*{Acknowledgements}
All experiments were performed using the Vermont Advanced Computing Center (VACC) 1400 processor cluster, an IBM e1350 High Performance Computing system.  This research was supported by a National Aeronautics and Space Administration (NASA) EPSCoR grant and National Science Foundation (NSF) Grant $\#$ DMS-$0940271$.  The original matlab code can be found at:
http://www.uvm.edu/$\sim$cdanfort/nolink/atmos/stalking-code.tar.  The authors would like to thank Kameron Harris, Nicholas Allgaier, and two anonymous reviewers for insightful comments and suggestions in preparing this manuscript.\par 

\bibliographystyle{elsarticle-num-names}
\bibliography{biblio}
\end{document}